\documentclass[runningheads]{llncs}
\usepackage{graphicx}
\usepackage{hyperref}

\usepackage{tikz}
\usetikzlibrary{arrows,automata,positioning}

\usepackage{stmaryrd}

\usepackage{prooftree}

\usepackage{amsmath}

\usepackage{comment}

\usepackage[frozencache,cachedir=.]{minted}

\usepackage{listings}

\usepackage{cleveref}

\usepackage{bussproofs}

\usepackage{enumitem}

\usepackage[font=small,skip=0pt]{caption}

\setlistdepth{9}

\setlist[itemize,1]{label=\normalfont\bfseries --}
\setlist[itemize,2]{label=$\bullet$}
\setlist[itemize,3]{label=$\bullet$}
\setlist[itemize,4]{label=$\bullet$}
\setlist[itemize,5]{label=$\bullet$}
\setlist[itemize,6]{label=$\bullet$}
\setlist[itemize,7]{label=$\bullet$}
\setlist[itemize,8]{label=$\bullet$}
\setlist[itemize,9]{label=$\bullet$}

\renewlist{itemize}{itemize}{9}

\newenvironment{scprooftree}[1]%
  {\gdef\scalefactor{#1}\begin{center}\proofSkipAmount \leavevmode}%
  {\scalebox{\scalefactor}{\DisplayProof}\proofSkipAmount \end{center} } 

\usepackage[colorinlistoftodos,textsize=scriptsize]{todonotes}

\newcommand{\coq}[1]{\lstinline[basicstyle=\normalsize\ttfamily]{#1}}

\begin{document}
\title{A Logic for Veracity: Development and Implementation}
%
%
\author{Daniel Britten\orcidID{0000-0002-7860-3595} \and
Steve Reeves\orcidID{0000-0002-3840-6060}}
\authorrunning{Britten and Reeves}
%
\institute{Department of Software Engineering, University of Waikato, Hamilton, New Zealand
\email{db130@students.waikato.ac.nz}\\\email{steve.reeves@waikato.ac.nz}}
\maketitle              
%


\begin{abstract}
In the business rules of supply chains, there are concerns around trust, truth, demonstrability and authenticity. These concerns are gathered together under the name ``veracity". 

In the work for this paper we were originally motivated by the requirement around organic certification in the wine industry in New Zealand, but veracity arises in many different situations and our formalisation shows how formal methods can give insights into many such practical problems.

One activity for formal methods involves taking informal processes and formalising them and subsequently building tools to support this formalisation and therefore the original processes too, and the work reported here is an example of that.

Here, then, we explore the idea of veracity in this spirit, give highlights of the development of a logic for it and show how that logic can be implemented in Coq, both for proof support and automation. 


\end{abstract}
\section{Introduction}

In supply chains of all sorts, there are concerns around trust, truth, demonstrability and authenticity. These arise because different people taking part in the chain, the actors, may not know each other or directly communicate, may have to use third parties (certainly for the software systems that they probably rely on), may have to take people or documents ``at their word" when they would actually like some more tangible evidence to go on, and so on.

Such supply chains may be for physical goods, or may be for software in some distributed development environment (say, using a Git \cite{Git24} repository for collaboration and the idea of a Software Bill Of Materials, SBOM \cite{SBOM22}, a counterpart of the established ``bill of materials" that accompanies physical goods as they move around).

The work in this paper was initially motivated by consideration of the supply chain ``problem" as part of the Veracity Project (which has a public face via The Veracity Lab \cite{VP24}). In that project some extensive use cases around the production and certification of organic wine in New Zealand were developed \cite{PeC22}.\footnote{New Zealand is one of the last countries in the world to put organic certification on a statutory basis, so this has also been a great opportunity to directly influence the government in select committees and with direct engagement with officials and ministers.}

There is a set of rules for certifying organic wine and these can be viewed as business rules, perhaps, but certainly the veracity of their application is an issue. Two questions with such sets of rules arise: are the rules sound?; how do we check that someone has followed the rules?

Our way of at least providing a basis for answering these questions is to develop a logic for veracity, and subsequently to mechanise it and make it useable ``under the hood" for legislators and producers.


This paper is about a rather new area for formal methods to target in such a fundamental way. Many formal methods techniques, of course, exist to deal with the various parts of a supply chain where systems and software are concerned. But we believe that this is the first work to try to formalise and then mechanise the underlying logic for such important and all-pervasive systems.

\subsection[Background to the Concept of ``Veracity"]{Background to the Concept of ``veracity"\footnote{Sections 1, 2 and 3 are (abbreviated) from the unpublished  \cite{Ree24}}}

When a piece of information is put out into the world it gets subjected to many attempts, both accidental and deliberate, to degrade it or tamper with it. When we are dealing with precious information, that is information which has value (cultural, monetary, scientific etc.), then having assurance that the information is always reliable is vital. When that information is not kept hidden or otherwise protected then this becomes a very hard problem. It may even be insoluble. This is the essence of the \emph{veracity problem}.

\emph{Veracity} seems to be a term that is widely used, but it is also hard to pin-down its meaning. In this paper we shall take it to mean, reflecting the concerns in the previous paragraph, that we have an assurance that the information has stayed constant. So, we say \emph{a piece of information has veracity} when we can check that it has not changed.

\subsection{Aims}


There is a long, deep, rich heritage of study of the concepts of truth and trust in many settings and many of them are formal, and very complicated and subtle. We want to start from scratch, not in order to just do something different, but in order to be able treat well, but lightly, those parts of veracity which can adequately be treated that way for our purposes (so, trust will be so treated). And then other parts (demonstrability, truth) will be looked at in more depth simply because they have not (as far as a literature search can show) been considered before, either with application in mind or in a formal way. Some examples that deal with adjacent ideas are \cite{artemov-fitting-book,ACGT21,taglia-22,roennedal-18,goranko-21,from-21,AMF20,michaelis-nipkow-17,renne-12,benthem-12,liu-lorini-17,ABE23,vBDE14}, though they all miss more or less of the components of veracity in some way.

  
To cut to the chase: we will look at intuitionistic logic since it seems to be clearly what is needed, as we argue below. None of the \emph{classical} modal logics work, for example, because of their classical basis, not because we do not like modal formalisms! The key to seeing this is that all those classical (and classical-including) logics lose information, which is precisely what a formalisation of veracity, as a starting point, must not do. Intuitionistic logic does not lose information, so \emph{obviously} it is the place to start. 

\subsection{Atomic Veracity}

Some statements have a sort of \emph{immediate} veracity, in the sense that they are newly minted by us and have not passed through any other hands.

Consider a couple of examples in more detail:
\begin{enumerate}
    \item 
    A bar code that we have ourselves just printed and associated with a physical object might be an example in a supply chain: this act might generate the information that \emph{this} bar code is stuck to  \emph{this} object that was produced by \emph{this} person, at \emph{this} time and \emph{this} place, has \emph{these} characteristics (composition, mass, etc.). We might say the information attests that ``this bar code really does identify this object";
    
    \item
    Or considering cultural objects, it might be an audio recording of a person giving their whakapapa (the Māori word that is usually condensed into the English word \emph{genealogy} though it is a much richer concept) together with its meta-data that we have just ourselves recorded and catalogued. Here the information is attesting to the association between the meta-data and the audio data.
\end{enumerate}

  These cases in some sense wear their veracity on their sleeve: it is immediate, we have ``a piece of veracity", the information that \emph{this} piece of data correctly describes \emph{this} object since I, at just this moment, made the association. This is our \emph{atomic veracity}. The claim or statement cannot be further analysed in terms of asking whose hands it has passed through, how it has be modified or added to since none of this has ever happened to it. 
  
  We might say (using logical terminology) that the piece of verifying information is a \emph{witness, proof,  testimony, piece of evidence} to the act of association, or the act of supporting a claim. The witness and the claim together are our judgements. It is these that we want to be able to objectify and then track. This track will be what we look to when someone says ``how do we know that this bar code correctly identifies this object?". 
  
  
  We might want to view this evidential information in more detail though. 
  A witness might include information of who $p$, where $\ell$, when $t$, how $m$ etc. is was made. In this case, the witness would not be an atomic name but an ``atomic" term. I.e. we might view a witness either as the atomic name $w$ or the atomic term $w(p,\ell,t,m)$. So a witness, a piece of evidence, might contain a lot of information, but from the point of view of the logic it is not further analysable. 

  One very important use for such a non-atomic a witness is that it contains the \emph{provenance} that contributes to the witness for a claim. This is often the meta-data attached to a digital artefact, for example.

How the data about it ``sticks" to the artefact (the bar code on the car part, the meta-data to the whakapapa audio file) is not what we are concerned about here. It is another technological problem that is being worked on and is outside our scope. So, we are  assuming it is possible and has been, or will be, done\footnote{This might be wishful thinking, but it has such big stakes for such large companies that we can assume it will happen one day, and indeed we now have cameras which help prove the authenticity of photos \cite{Reu23}, or the use of isotopes to track the provenance of food and prove its authenticity \cite{Iso23}, or very, very close-up photos of objects from particular angles to support authenticity \cite{Ogr24}, and so on. In any case, veracity is still an interesting idea to try to reason about.}.


\section{A Logic for Veracity}

We let letters like $A$, $B$, etc. stand for a \emph{claim} of veracity (for example ``This wine is organic", ``This crankshaft is a genuine BMW one", ``This whakapapa follows the correct line for relating her ancestors"), which is a form of proposition that \emph{holds, is supported etc.} when the veracity claimed is appropriately \emph{witnessed}, upheld by data, by a person's statement, by direct knowledge, evidence, that the thing is what we say it is, came from where we say it came from, was grown as we say it was grown, etc.

Then a $judgement$ $a \in A$ is a \emph{veracity judgement}
that $A$ has witness $a$. A judgement like this is upheld, or perhaps we might say that $A$ has veracity because it is witnessed by $a$, when this judgement appears as the root of a proof tree constructed according to the rules that follow. There are other forms of judgement too: for example the judgement that $A$ is a claim (of veracity) is $A\ veracity\ claim$. We will see other forms (to do with equality or computation) later, in addition to these two.

To make the idea of judgement clearer (we hope), since it is not familiar even to most (formal) logicians, let alone others (though computing people use them all the time, formally, in \emph{type declarations} like $x : Int$) we here borrow, paraphrase and present a diagram from \cite{Lof84}, page three:


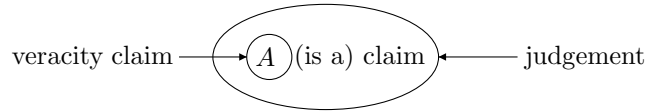
\begin{center}
\begin{tikzpicture}[scale=0.1]
\tikzstyle{every node}+=[inner sep=0pt]
\draw [black] (9.5,-14.4) ellipse (15 and 7);
\draw (14,-14.4) node {(is a) claim};
\draw [black] (2,-14.4) circle (3);
\draw (1.5,-14.4) node {$A$};
\draw [black] (-10,-14.4) -- (-1,-14.4);
\fill [black] (-1,-14.4) -- (-2,-13.9) -- (-2,-14.9);
\draw (-21.5,-14.4) node {veracity claim};
\draw [black] (24.5,-14.4) -- (35,-14.4);
\fill [black] (24.5,-14.4) -- (25.5,-13.9) -- (25.5,-14.9);
\draw (44,-14.4) node {judgement};
\end{tikzpicture}
\end{center}

\noindent and when showing the witness that upholds the veracity claim we have the form

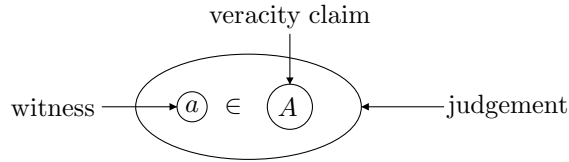
\begin{center}
\begin{tikzpicture}[scale=0.1]
\tikzstyle{every node}+=[inner sep=0pt]
\draw [black] (9.5,-14.4) ellipse (15 and 7);
\draw [black] (2,-14.4) circle (2);
\draw [black] (15,-14.4) circle (3);
\draw (2,-14.4) node {$a$};
\draw (14.7,-14.4) node {$A$};
\draw (7.5,-14.4) node {$\in$};
\draw [black] (-10,-14.4) -- (0,-14.4);
\fill [black] (0,-14.4) -- (-1,-13.9) -- (-1,-14.9);
\draw (-16.5,-14.4) node {witness};
\draw [black] (15,-4.7) -- (15,-11.4);
\fill [black] (15,-11.4) -- (14.5,-10.5) -- (15.5,-10.5);
\draw (15,-2.4) node {veracity claim};
\draw [black] (24.5,-14.4) -- (35.5,-14.4);
\fill [black] (24.5,-14.4) -- (25.5,-13.9) -- (25.5,-14.9);
\draw (44,-14.4) node {judgement};
\end{tikzpicture}
\end{center}

\noindent and note that, as implied by the picture, the $\in$ symbol is part of the judgement language, not part of the witness or claim languages.

\subsection{No Veracity}

There is a special veracity claim $\bot$ which has no witnesses, i.e. it is the claim that never has veracity, and a judgement that makes a claim about it can never be upheld.

This leads to our first proof rule: 

$$
\begin{prooftree}
a \in \bot
\justifies
a \in A
\using
{\bot^-}
\end{prooftree}
$$

This rule says that if you, in the course of your reasoning, somehow have shown that the claim $\bot$ (that can never have veracity) does in fact have it, then you can show that $anything$ has veracity. We call this rule $\bot^-$ for ``$\bot$ elimination".

Martin-L\"of makes this comment on his corresponding rule \cite{Lof84}: 

$$
\begin{prooftree}
c \in N_0
\justifies
R_0(c) \in C(c)
\using
{N_0\ elimination}
\end{prooftree}
$$

``The explanation of this rule is that we understand that we shall never get an element $c \in N_0$, so that we shall never have to execute $R_0(c)$. Thus the set of instructions for executing a program of the form $R_0(c)$ is vacuous. It is similar to the programming statement \emph{abort} introduced by Dijkstra."

We use the same justification and motivation for our rule too.

\subsection{Adding Claims Together}
\label{sec:adding}

$$
\begin{prooftree}
a \in A \quad b \in B
\justifies
(a,b) \in A \land B
\using
{\land^+}
\end{prooftree}
$$

$$
\begin{prooftree}
(a,b) \in A \land B
\justifies
a \in A
\using
{\land^-1}
\end{prooftree}
$$

$$
\begin{prooftree}
(a,b) \in A \land B
\justifies
b \in B
\using
{\land^-2}
\end{prooftree}
$$

Here we are formalising the idea that if two veracity claims $A$ and $B$ are witnessed then the combined claim that $A$ together with $B$ has veracity is also witnessed, and that witness we choose to denote by the pairing of the component witnesses.

Note that this is a simple use of the the idea also of information being preserved around claims and their witnesses even when they are composed together.

\subsection{Choice Between Claims}\label{sec:choice}

One immediate place where this information preservation becomes perhaps a little unfamiliar is when we try to think about what saying ``we have claims $A$ and $B$ and we know that they each have a witness, so we know that one or the other has one: that is, a claim of $A$ or $B$ is witnessed". We might choose to formalise this by saying

$$
\begin{prooftree}
a \in A \quad b \in B
\justifies
a \in A \lor B
\using
{}
\end{prooftree}
$$

The point here is that (first) this rule has exactly the same premises as the one above, and avoiding such points of choice amongst rules is generally (for coherence) a good thing. But more importantly (at the formalisation level) is that we have lost information here. The conclusion does not record which of the alternatives we have relied on to reach it: did we justify the claim of one or the other because of the first witness, or the second?

Righting these two points means doing something like

$$
\begin{prooftree}
a \in A 
\justifies
i(a) \in A \lor B
\using
{\lor^+1}
\end{prooftree}
$$

$$
\begin{prooftree}
b \in B
\justifies
j(b) \in A \lor B
\using
{\lor^+2}
\end{prooftree}
$$

So, if we have a witness to a claim of $A$ then we certainly have a witness to a claim of either $A$ or $B$, and we ``tag" the witness in the conclusion so that we do not lose the information about which claim the claim of one or the other relies on.

Now consider the case where we know that a certain witness $c$ upholds the claim that $A \lor B$. What can we deduce, if anything, from this? 

First note that our two introduction rules mean that a witness to a claim like this must in fact have a tag since tags are introduced by the only rule that could have allowed us to deduce the claim of $A \lor B$. So, we have a case analysis to do: if the witness to this composite claim is tagged with $i$ then we know it is $A$ that we relied on and similarly with $j$ and $B$. This preservation of all the information allows us to dismantle the composite claim:

$$
\begin{prooftree}
i(a) \in A \lor B
\justifies
a \in A 
\using
{\lor^-1}
\end{prooftree}
$$

$$
\begin{prooftree}
j(b) \in A \lor B
\justifies
b \in B 
\using
{\lor^-2}
\end{prooftree}
$$




\subsection{Veracity Claims with Prior Assumptions}
\label{sec:implication}

Imagine that by assuming that claim $A$ has veracity, i.e. that the judgement $x \in A$ has been shown for some arbitrary witness $x$,  we can show that claim $B$ has veracity, i.e. we can show $b \in B$.
  
Denote this state of affairs by \emph{a claim that depends on an assumption}:
  
  $$
  x \in A \vdash b \in B
  $$
  
  The $\vdash$ is a turnstile (because of its shape) and is a relation between judgements. As we will see, it relates a set of judgements (the assumptions) with a single judgement (the conclusion). We will call this generalised form a judgement too (since the conclusion is certainly a judgement, and it's usually the focus of our concerns).
  
  As is typical, we introduce an $implication\ claim$ to reflect this, i.e. to discharge the assumption, so the claim becomes
  
  $$
  A \rightarrow B
  $$ 
  
\noindent but what would a witness to $this$ claim plausibly look like?
 
Given any witness $x$ to the claim $A$ then it is possible to \emph{construct} a witness for the claim $B$. That is, there is a function which given any witness to $A$ will compute a witness to $B$, so

$$
\lambda b \in A \rightarrow B
$$

 The witness to an implicative claim like $A \rightarrow B$ should be a function that takes a witness to the claim $A$ and turns it into a witness for the claim $B$.\footnote{For expression $e$ and variable $x$, the expression $(x)e$ is an expression where all free occurrences of $x$ in $e$ become bound by this $(x)$. The expression $(x)e$ called an abstraction (of $e$ by $x$). For expression $(x)e$ and expression $a$, $(x)e(a)$ is an expression where all occurrences of $x$ in $e$ bound by this $(x)$ are replaced by $a$. The expression $(x)e(a)$  is called the application of $(x)e$ to $a$. Note that $b$ in the example must be an abstraction for this judgement to be well-formed.}

 In general, this allows us to build a function that, given a whole set of basic veracity claims and their witnesses (the assumptions), builds for us a witness for a complex veracity claim. This function can then be read as a process to be followed which, given starting veracity claims, will assure that a complex veracity claim can be successfully and correctly made. 

Implication allows us to define negation in terms of $\bot$: $\lnot A$ is $A \rightarrow \bot$. A witness to a claim of $\lnot A$ takes a witness to $A$ and gives us a witness to $\bot$. But $\bot$ has no witness, so a witness to $A$ is not possible, as expected by our informal understanding of saying a claim has no witnesses, i.e. that $\lnot A$ holds.



The requirement that to justify a disjunction of claims it has to be demonstrated which of the claims were justified before (which is the role that the tags on the witnesses are playing in the rules) means that, for example, the claim $A \lor \lnot A$ is also not justifiable without saying which claim is witnessed: $A \lor \lnot A$ doesn't survive the question: yes, but can you show, whatever $A$ is, the witness that assures the veracity of the claim here?
  
And the view that witnesses to implications are functions leads us in the same direction: to the thought that this is reinterpreting  intuitionistic logic


Note: in fact these rules that we have given in the narrative above (in order to motivate the whole idea) are not general enough. In particular, the elimination rules need to introduce non-canonical forms of expression, and then we need equality rules to ``compute" the value (i.e. the canonical form) of a non-canonical form. The full logic has this, but it turns out that for the examples we were motivated by in the supply chain rules need only introduction rules, and that is what the examples later and the automation currently concentrates on. The full rules (that go beyond those needed for the supply chain examples) are in \Cref{sec:thelogic}. And as an example of an even more general rule (which the given elimination rules are special cases of, and where we have to use the full logic of dependent claims to prove them, in fact) see  \Cref{app:choice}. 

\section{More Actors}

The argument so far is that the logic work above covers the demonstrability (checking a proof is easy) and truth aspects of veracity. What is not yet settled is the trust aspect (the authenticity is left for now--yet to have any ideas on how it might be treated, or even what it is), and once we start to think about trust, we think about people and the relationships between them.

The section above works well when one person is collecting and making veracity claims. It is a one-person logic because we never mention who is making claims, so we cannot tell how many people might be, so we can only correctly assume it is one person from the form of the rules. In other words, there are no rules for combining or tracking veracity claims made by several actors.

One way to perhaps tackle this is to add a name (of an actor) to each justified judgement. So, if actors $k$ and $\ell$ from a set $Act$ have made claims then we might have two judgements $a^k \in A$ and $b^\ell \in B$, that is actor $k$ has made claim $A$ with witness $a$, and similarly for $\ell$, $b$ and $B$.

This now adds a second dimension to our logic above. The first dimension dealt with one actor, so we can think of all the judgements before as being abbreviations (because there's only one actor $k$) of judgements of the form $a^k \in A$, so we left the $k$ out because it never varied. Now the second dimension is around how actors become incorporated into the logic.

\subsection{Relating Actors}

Having introduced more than one actor we now need to think about how, from a veracity point of view, they can be related. 

Keeping to the idea that we think of simple cases to guide us rather than trying to do everything we might wish all at once, the question: what relationship between actors is a useful one (there will be others) to consider? Fundamentally, surely, is one of trust: does this actor trust that actor? Once we know who trusts who we can plausibly expect things like $k$ trusting $\ell$ mean that any judgement that $\ell$ has accepted allows $k$ accept that judgement. So, roughly, we would say that $\vdash a^\ell \in A$ leads to $\vdash a^k \in A$ if $k$ trusts $\ell$. If we denote the trust relation by $T \subseteq Act \times Act$ then $k$ trusts $\ell$ will be $kTl$. We propose a rule

$$
\begin{prooftree}
a^\ell \in A \quad kTl
\justifies
a^k \in A 
\using
{trust\ T}
\end{prooftree}
$$

and we can picture the relation as

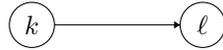
\begin{center}
\begin{tikzpicture}[scale=0.1]
\tikzstyle{every node}+=[inner sep=0pt]
\draw [black] (9.5,-14.4) circle (3);
\draw (9.5,-14.4) node {$k$};
\draw [black] (32.2,-14.4) circle (3);
\draw (32.2,-14.4) node {$\ell$};
\draw [black] (12.5,-14.4) -- (29.2,-14.4);
\fill [black] (29.2,-14.4) -- (28.4,-13.9) -- (28.4,-14.9);
\end{tikzpicture}
\end{center}





A note: in the rule $trust\ T$, the $kTl$ premise should be considered to be syntactic sugar intended to focus on one particular element of $T$ for the purposes of clarity, since the presence of $T$ in the name of the rule is enough to justify the inference given that the remaining premise and the conclusion show what must be checked (that indeed $kTl$) and that can be done by reviewing the definition of $T$. This means that a proof always takes place in the context of one or more trust relations, and that the name of the rule is actually a name $trust$ and an abbreviated proviso, namely $the\ required\ actors\ must\ be\ related\ to\ one\ another\ in\ an\ in-context\ trust\ relation$. $k$ and $\ell$ in the rule can be identified by noting that $k$ is the actor mentioned in the conclusion and $\ell$ the actor mentioned in the premise.

\subsection{Trust Relations}

We can explore, even with this simple basis, how veracity works. 



Given $T$ as

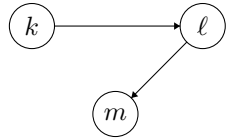
\begin{center}
\begin{tikzpicture}[scale=0.1]
\tikzstyle{every node}+=[inner sep=0pt]
\draw [black] (9.5,-14.4) circle (3);
\draw (9.5,-14.4) node {$k$};
\draw [black] (32.2,-14.4) circle (3);
\draw (32.2,-14.4) node {$\ell$};
\draw [black] (20.6,-26.1) circle (3);
\draw (20.6,-26.1) node {$m$};
\draw [black] (12.5,-14.4) -- (29.2,-14.4);
\fill [black] (29.2,-14.4) -- (28.4,-13.9) -- (28.4,-14.9);
\draw (20.85,-13.9) node [above] {};
\draw [black] (30.09,-16.53) -- (22.71,-23.97);
\fill [black] (22.71,-23.97) -- (23.63,-23.75) -- (22.92,-23.05);
\draw (25.88,-18.77) node [left] {};
\end{tikzpicture}
\end{center}

we have

$$
\begin{prooftree}
\[a^m \in A \quad lTm
\justifies
a^\ell \in A 
\using
{trust\ T} \] \quad kTl
\justifies
a^k \in A 
\using
{trust\ T}
\end{prooftree}
$$

Given this example we might ask: can $any$ binary relation between actors be a trust one? No; it surely needs to be at least reflexive and certainly not symmetric: we trust ourselves, and if we trust someone does it follow that they should trust us? 

Note that the proof above seems to show that trust is also transitive: it turns out to be a property of our simple rule. Does that call the simple rule into question, since it a stretch to accept that if I trust someone, and they trust someone else, then I should trust that someone else? 

Well, we make the point that trust here is ``100\% trust" which explains this rule and how transitivity emerges in this pointwise way. We will return to this below.

Another derivable rule which seems to be a good thing: if two people see veracity in two different things and one trusts the other then the first person believes the conjunction.


Here is a derivation (a proof tree) that shows the validity of this derived rule:

$$
\begin{prooftree}
\[a^k \in A \quad kTl
\justifies
a^\ell \in A 
\using
{trust\ T} \] \quad b^\ell \in B
\justifies
(a,b)^\ell \in A \land B
\using
{\land^+}
\end{prooftree}
$$

\noindent and where we have generalised (not the final generalisation step, as we will see) the $\land^+$ rule from Section \ref{sec:choice} by adding actors.



\subsection{Degrees of Trust}

This brings the final augmentation, that we need \emph{degrees of trust} to make things work. We write

$$
a_{0.5}^k \in A
$$ 

\noindent for $k$ believes with strength 0.5 that $a$ supports the claim $A$ (and we drop the subscript in the case it's 1.0).

Then the apparent transitivity above only works if $kT_{1.0}\ell$ and $lT_{1.0}m$, i.e. $k$ trusts $\ell$ completely, and the same for $\ell$ and $m$, i.e.
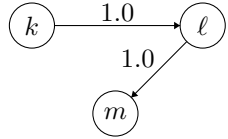
\begin{center}
\begin{tikzpicture}[scale=0.1]
\tikzstyle{every node}+=[inner sep=0pt]
\draw [black] (9.5,-14.4) circle (3);
\draw (9.5,-14.4) node {$k$};
\draw [black] (32.2,-14.4) circle (3);
\draw (32.2,-14.4) node {$\ell$};
\draw [black] (20.6,-26.1) circle (3);
\draw (20.6,-26.1) node {$m$};
\draw [black] (12.5,-14.4) -- (29.2,-14.4);
\fill [black] (29.2,-14.4) -- (28.4,-13.9) -- (28.4,-14.9);
\draw (20.85,-13.9) node [above] {$1.0$};
\draw [black] (30.09,-16.53) -- (22.71,-23.97);
\fill [black] (22.71,-23.97) -- (23.63,-23.75) -- (22.92,-23.05);
\draw (25.88,-18.77) node [left] {$1.0$};
\end{tikzpicture}
\end{center}

and that makes the apparent transitivity look reasonable.

So, we recast the $trust\ T$ rule as
  
$$
\begin{prooftree}
kT_xl \quad a_y^\ell \in A
\justifies
a_{x.y}^k \in A
\using
{trust\ T}
\end{prooftree}
$$

If instead $kT_{0.5}\ell$ and $lT_{0.4}m$ then I would say $kT_{0.2}m$ and the proof above supports this, rewritten as

$$
\begin{prooftree}
 kT_{0.5}\ell
 \[ lT_{0.4}m \quad a^m \in A 
\justifies
a_{0.4}^\ell \in A 
\using
{trust\ T} \]
\justifies
a_{0.2}^k \in A 
\using
{trust\ T}
\end{prooftree}
$$

i.e. if $a_{1.0}^m \in A$ and $kT_{0.5}\ell$ and $lT_{0.4}m$ 

\begin{center}
\begin{tikzpicture}[scale=0.1]
\tikzstyle{every node}+=[inner sep=0pt]
\draw [black] (9.5,-14.4) circle (3);
\draw (9.5,-14.4) node {$k$};
\draw [black] (32.2,-14.4) circle (3);
\draw (32.2,-14.4) node {$\ell$};
\draw [black] (20.6,-26.1) circle (3);
\draw (20.6,-26.1) node {$m$};
\draw [black] (12.5,-14.4) -- (29.2,-14.4);
\fill [black] (29.2,-14.4) -- (28.4,-13.9) -- (28.4,-14.9);
\draw (20.85,-13.9) node [above] {$0.5$};
\draw [black] (30.09,-16.53) -- (22.71,-23.97);
\fill [black] (22.71,-23.97) -- (23.63,-23.75) -- (22.92,-23.05);
\draw (25.88,-18.77) node [left] {$0.4$};
\end{tikzpicture}
\end{center}

then $a_{0.2}^k \in A$. 


\subsection{The Full Logic}

For reasons of space, we have barely scratched the surface of the complete logic, but have sought to give motivation to the idea and then a few formal highlights. The complete logic can be summarised as: Martin-L\"of's intuitionistic type theory, with the trust rule, actors and weights added, and then the whole reinterpreted as we have shown above. In particular, where the original logic talked about proof objects (or programs) and propositions (or types, specifications) we talk about witnesses and claims, respectively. We also rest on Martin-L\"of's soundness argument for the logic \cite{Lof84} (though we have also shown the rules are sound more conventionally via a semantics that rests on defining the meaning of a claim as being the set of all its witnesses).

\section{Mechanisation via Coq}

We now pass to the next stage of this work, that of mechanising the logic, via Coq, in order that we can take advantage of the proof assistant to both help in finding and then checking our proofs. Given that ultimately we want tools which will help non-logicians (or perhaps even non-formal methods people)  use the logic to explore veracity concerns, this is a vital step.

Coq~\cite{coq} is a proof assistant with a dependently typed programming language called Gallina. The mechanisation of the veracity logic in Coq enforces the rules of the logic and facilitates constructing correct proof trees, both interactively and semi-automatically. The approach to the mechanisation centres around defining the dependent inductive type \coq{proofTreeOf Ps j} which is the type for correct proof trees, i.e. proof trees that have this judgement at their root, of the judgement \coq{j}, with the assumptions \coq{Ps}. This approach makes interactive proofs using Coq's proof mode possible, where the user is actually constructing a value of type \coq{proofTreeOf Ps j} rather than proving a typical Coq proposition. It also makes it possible to use regular functional programming techniques on proof trees. This has been used to generate \LaTeX\ renderings of proof trees as well as simple natural language renderings of proof trees. The automation, much of which is future work, also relies on being able to construct proof trees in a similar way to how trees can be constructed in other functional programming languages.


\vspace{-0.3cm}

\subsection{The Proof Tree Type}
The type for proof trees constrains them to be correct proof trees of the root judgement, however the type does not constrain which assumptions can be made.

\begin{listing}
  \caption{Selected rules from proof tree type}
  \label{list:proofTreeOf}
\begin{minted}[fontsize=\small, breaklines]{coq}
Inductive proofTreeOf : list judgement -> judgement -> Type :=
  | assume e a c
    : proofTreeOf [((AtomicEvid e) \by a \in c)]
                  ((AtomicEvid e) \by a \in c)
  | and_intro e1 e2 a C1 C2 Ps Qs Rs
      (H : (Ps ++ Qs ==? Rs) = true)
      (L: proofTreeOf Ps (e1 \by a \in C1))
      (R: proofTreeOf Qs (e2 \by a \in C2))
    : proofTreeOf Rs ({{e1, e2}} \by a \in (C1 /\' C2))
  | or_elim1 e1 a C1 C2 Ps
      (M: proofTreeOf Ps ((Left e1) \by a \in (C1 \/' C2)))
    : proofTreeOf Ps (e1 \by a \in C1)
  | impl_elim x bx y a C1 C2 Ps Qs Rs
      (H1 : (Ps ++ Qs ==? Rs) = true)
      (H2 : notUsedInInnerLambda x bx = true)                
      (L: proofTreeOf Ps ((Lambda x bx) \by a \in (Implies C1 C2)))
      (R: proofTreeOf Qs (y \by a \in C1))
    : proofTreeOf Rs ((apply_lambda x bx y H2) \by a \in C2)
\end{minted}
\vspace{-0.3cm}
\end{listing}

The intention is that when carrying out a proof either interactively or semi-automatically, the person constructing the proof will make use of only the assumptions and trust relations that are valid in their use-case. It is still possible to programmatically analyse a proof tree after it has been constructed to ensure that only certain assumptions and trust relations have been used.

In \Cref{list:proofTreeOf} we demonstrate how the \coq{proofTreeOf Ps j} type is defined, focusing on a selected rules. The full definition is available on GitHub\footnote{\url{https://github.com/Coda-Coda/Veracity-Logic-Mechanised/tree/FM-24}}.

\begin{itemize}
  \item The base case is handled by the rule \coq{assume} which allows us to assume that actor \coq{a} holds claim \coq{C} by the evidence \coq{e} provided. This rule should only be used when it is appropriate to do so given the use-case.
  \item The rule \coq{and_intro} requires a proof tree showing that claim \coq{C1} is held by actor \coq{a} with the evidence \coq{e1} as well as a proof tree that claim \coq{C2} is also held by actor \coq{a} with the evidence \coq{e2}. Given this, we have a proof tree that the conjunction of claims \coq{C1} and \coq{C2} is held by actor \coq{a} with the evidence \coq{\{\{e1, e2\}\}}. This corresponds to the rule labelled ``$\wedge^{+}$'' in \Cref{sec:adding}.
  \item The rule \coq{or_elim1} captures the rule labelled ``$\vee^{-}1$'' in \Cref{sec:choice}.
  \item The rule \coq{impl_elim} allows the application of the function that is included in the evidence for a proof of \coq{C1} to \coq{C2}, to potentially different evidence of \coq{C2}. Such a function is referred to as $b$ in \Cref{sec:implication}.
  
  
\end{itemize}

\subsection{Interactive Proof Tree Construction}
To construct a proof tree interactively, we can use Coq tactics as in \Cref{list:processExample}. As shown, the proof follows from the application of the \coq{impl_intro} and \coq{and_intro} rules, followed by assuming $C_1$, $C_2$ and $C_3$ involving the evidence $\ell$, $s$ and $c$, respectively.
This is an example from a case study, as discussed in \Cref{app:caseStudyExamples}.

\begin{listing}
  \caption{Example proof script}
  \label{list:processExample}
\begin{minted}[fontsize=\small, breaklines]{coq}
Definition j1 := x \by P \in c1. Definition j2 := y \by P \in c2.
Definition j3 := z \by P \in c3.
Definition process_example 
  : proofTreeOf_wrapped P (c3 ->' (c2 ->' (c1 ->' (c1 /\' c2 /\' c3)))).
Proof. eexists _ _.
eapply impl_intro with (x:=_z_) (Ps := [j3]) (Qs:=[]). 1-3: shelve.
eapply impl_intro with (x:=_y_) (Ps := [j2;j3]) (Qs:=[j3]). 1-3: shelve.
eapply impl_intro with (x:=_x_) (Ps := [j1;j2;j3]) (Qs:=[j2;j3]).
  1-3: shelve.
eapply and_intro with (Ps := [j1;j2]). shelve. eapply and_intro. shelve.
apply assume with (e := _x_). apply assume with (e := _y_).
apply assume with (e := _z_).
Unshelve. all: reflexivity. Defined.
\end{minted}
\vspace{-0.3cm}
\end{listing}

\subsection{Generating Renderings of Proof Trees}
Once we have constructed proof trees we can use standard functional programming techniques to convert value of the \coq{proofTreeOf Ps j} type to strings such as renderings in \LaTeX\ or natural language. The code that implements this is omitted for brevity, but is available on GitHub\footnote{\url{https://github.com/Coda-Coda/Veracity-Logic-Mechanised/tree/FM-24}}. In \Cref{fig:processExample}, we show the result of rendering the proof from \Cref{list:processExample}. A natural language rendering of the same proof is included in \Cref{app:naturalLanguage}. Both renderings are computed by Coq from a completed proof, a value of type \coq{proofTreeOf Ps j}. The notation $P$, as a superscript to evidence, applies to all parts of the witness.


\vspace{-0.3cm}
\begin{figure}
  \begin{scprooftree}{0.8}\AxiomC{$ C_{1} \textit{ is a veracity claim} $} \RightLabel{ $ assume $}\UnaryInfC{$ x^{P} \in C_{1} \vdash_{} x^{P} \in C_{1} $}\AxiomC{$ C_{2} \textit{ is a veracity claim} $} \RightLabel{ $ assume $}\UnaryInfC{$ y^{P} \in C_{2} \vdash_{} y^{P} \in C_{2} $} \RightLabel{ $ \wedge^{+} $} \BinaryInfC{$ x^{P} \in C_{1}, y^{P} \in C_{2} \vdash_{} (x, y)^{P} \in (C_{1} \wedge C_{2}) $}\AxiomC{$ C_{3} \textit{ is a veracity claim} $} \RightLabel{ $ assume $}\UnaryInfC{$ z^{P} \in C_{3} \vdash_{} z^{P} \in C_{3} $} \RightLabel{ $ \wedge^{+} $} \BinaryInfC{$ x^{P} \in C_{1}, y^{P} \in C_{2}, z^{P} \in C_{3} \vdash_{} ((x, y), z)^{P} \in ((C_{1} \wedge C_{2}) \wedge C_{3}) $} \RightLabel{ $ \rightarrow^+ $} \UnaryInfC{$ y^{P} \in C_{2}, z^{P} \in C_{3} \vdash_{} \lambda(x)(((x, y), z))^{P} \in (C_{1} \rightarrow ((C_{1} \wedge C_{2}) \wedge C_{3})) $} \RightLabel{ $ \rightarrow^+ $} \UnaryInfC{$ z^{P} \in C_{3} \vdash_{} \lambda(y)(\lambda(x)(((x, y), z)))^{P} \in (C_{2} \rightarrow (C_{1} \rightarrow ((C_{1} \wedge C_{2}) \wedge C_{3}))) $} \RightLabel{ $ \rightarrow^+ $} \UnaryInfC{$ \lambda(z)(\lambda(y)(\lambda(x)(((x, y), z))))^{P} \in (C_{3} \rightarrow (C_{2} \rightarrow (C_{1} \rightarrow ((C_{1} \wedge C_{2}) \wedge C_{3})))) $}\end{scprooftree}
  \caption[position=top]{Generated \LaTeX\ rendering of the proof constructed in \Cref{list:processExample}}
  \label{fig:processExample}
\end{figure}

\vspace{-1.2cm}

\subsection{Automating Proof Tree Construction}
\label{sec:automation}

This section gives an insight into a technique for automating proof tree constructions using a much simplified version of the \coq{proofTreeOf} type. Bringing this level of automation to the version of \coq{proofTreeOf Ps j} described previously is left for future work.

Rather than incorporating evidence into the type \coq{proofTreeOf}, we instead only include the actor and claim as a part of \coq{j} here and omit the list of assumptions. Then we define a Coq function, \coq{computeEvidence}, to compute the evidence for each line in the proof tree. A similar approach is taken for the list of assumptions.

% ...


The main automation approach taken is to define a function which takes a proof tree with holes and fills in all the holes by one level deeper.
This function is then repeated multiple times up until a depth limit is reached.
For each unique proof goal and use-case, the user must define the function that takes the proof tree one level deeper. For an example of such a function see \Cref{list:proofStepExample}. This could be facilitated by a web page with a helpful user interface in the future. 

\vspace{-0.9cm}

\begin{listing}[h]
  \caption[position=top]{An example function for taking a proof tree one level deeper}
  \label{list:proofStepExample}
\begin{minted}[autogobble,fontsize=\small, breaklines]{coq}
Definition proofStepExample (j : judgementPart) : list (proofTreeOf j) :=
    match j with JudgementPart a c => 
      (if (a =? a1) && (c =? C) then [assume e a c] else []) ++
      (if (a =? a1) && (c =? (C /\' C)) then [assume e a c] else []) ++
      match c with
        | And C1 C2 => [and_intro a C1 C2 (hole _) (hole _)] 
        | _ => [] end end.
\end{minted}
\vspace{-0.3cm}
\end{listing}

 \vspace{-.4cm}

The user-defined function is then passed to the function shown in \Cref{list:oneLevelDeeper}, which applies the user-defined step function to a proof tree with holes and returns a list of proof trees that have had any holes filled in `one level deeper'. This function is then generalised to take in a list of proof trees with holes, making use of \coq{flat_map}. The resulting function, which both takes in and returns a list of proof trees with holes, is then repeated on its own output until the depth limit is reached. Typically, the initial list would be a list containing the single proof tree of the claim aiming to be proved with a hole as its evidence.

\ 

\vspace{-1.0cm}

\begin{listing}[h]
  \caption[position=top]{Part of the helper function for semi-automated proof search}
  \label{list:oneLevelDeeper}
\begin{minted}[fontsize=\small, breaklines]{coq}
Fixpoint oneLevelDeeper (step : forall j : judgementPart, list (proofTreeOf j)) (j : judgementPart) (p : proofTreeOf j)
  : list (proofTreeOf j) :=
  match p with
  | assume e a c => [(assume e a c)]
  | bot_elim a C M => map (bot_elim a C) (oneLevelDeeper step _ M)
  | and_intro a C1 C2 L R =>
        map (fun L2 => and_intro a C1 C2 L2 R) (oneLevelDeeper step _ L)
     ++ map (and_intro a C1 C2 L) (oneLevelDeeper step _ R)
\end{minted}
\vspace{-0.3cm}
\end{listing}

\vspace{-.3cm}

An important consideration with this approach to proof search is that in the context of these veracity proofs, it matters \emph{which} proof we have of a certain claim's veracity. Proof-irrelevance is not appropriate and so the usual techniques for proof search, which typically result in only one proof being found, are less useful. For example, there are four different proofs of the claim $C \wedge C \wedge C \wedge C$ generated using the step function shown in \Cref{list:proofStepExample}. It would not be sufficient to only generate one of these proofs of this illustrative claim automatically because in realistic examples what can be assumed is constrained. The resulting \LaTeX\ renderings of the four generated proofs are shown in \Cref{app:automation}.

\section{Future Work}
The mechanisation does not handle non-canonical evidence and the associated rules, such as given in \Cref{sec:thelogic} and \Cref{app:choice}. Future work would involve including this and a richer representation of claims, capturing the notion of families of sets of claims.
Incorporating weights for trust relations and throughout the mechanisation is also future work.

As described in \Cref{sec:automation}, the automation only works with a simplified set of incomplete rules from the logic. In particular, reasoning about `or' elimination and implications in a way which supports automation is not yet possible. This stems from removing evidence from the type of \coq{proofTreeOf} which makes it possible to have, for instance, lists of proofs of claims. With the \coq{proofTreeOf} depending on evidence, such lists would be ill-typed. While there are work-arounds, the natural approaches using \coq{flat_map}, for example, would need to be reworked in future work extending the automation to all the rules.

\section{Conclusions}

We started this work because others in the wider Veracity project needed some formulation for what the word ``veracity" might mean. The abstraction that the logic embodies gives insight into the common ways of reasoning in many particular example domains. And its formal nature gives a basis for computational tools to be developed. Thus, this work has been a classic exercising of the ``use case" for formal methods.

Considering the various goals we set ourselves, we summarise as follows: 
\begin{itemize}
    \item assuming we get true claims as our starting point (our assumptions and undischarged premisses), then \emph{truth} is maintained. Also \emph{trust} is expressed and handled in the rules. A proof tree \emph{demonstrates} all this. So, three of the four components of veracity are preserved in the logic;

    \item regarding the fourth component of veracity, \emph{authenticity}: at least one aspect of it may be expressed in witnesses (recall they can be load-carrying). We have found in our use of the logic that another aspect that points to problems with authenticity is when a proof can be constructed but with certain assumptions undischarged, since those assumptions are often failures of authenticity;

    \item the proof support from Coq, for the sorts of problems that arise in our examples around supply chains and organic certification, has been achieved, plus some user-friendly mechanisation to render proof trees in somewhat natural language which typical users can more readily relate to.
\end{itemize}


%
%
%
 \bibliographystyle{splncs04}
 \bibliography{veracity,refs,daniel}

\begin{thebibliography}{10}
\providecommand{\url}[1]{\texttt{#1}}
\providecommand{\urlprefix}{URL }
\providecommand{\doi}[1]{https://doi.org/#1}

\bibitem{Iso23}
Food authenticity and profiling, \url{https://www.thermofisher.com/nz/en/home/industrial/food-beverage/food-authenticity-labeling.html}

\bibitem{Git24}
Git version control \url{https://git-scm.com/}

\bibitem{VP24}
The veracity lab \url{https://veracity.wgtn.ac.nz/}

\bibitem{Reu23}
Reuters new proof of concept employs authentication system to securely capture, store and verify photographs (2023), \url{https://www.canon.co.uk/press-centre/press-releases/2023/08/reuters-new-proof-of-concept-employs-authentication-system/}

\bibitem{Ogr24}
Alberge, D.: Manufacturing giant develops revolutionary system to detect counterfeit art (2024), \url{https://www.theguardian.com/artanddesign/2024/jan/27/manufacturing-giant-develops-revolutionary-system-to-detect-counterfeit-art}

\bibitem{ACGT21}
Aldini, A., Curzi, G., Graziani, P., Tagliaferri, M.: Trust evidence logic. In: Vejnarov{\'a}, J., Wilson, N. (eds.) Symbolic and Quantitative Approaches to Reasoning with Uncertainty. pp. 575--589. Springer International Publishing, Cham (2021)

\bibitem{ABE23}
Alwhishi, G., Bentahar, J., Elwhishi, A.: Multi-valued model checking a smart glucose monitoring system with trust. In: 2023 International Wireless Communications and Mobile Computing (IWCMC). pp. 1697--1702 (2023). \doi{10.1109/IWCMC58020.2023.10183263}

\bibitem{AMF20}
Andrade~Guzman, J.M., Hernandez~Quiroz, F.: Natural deduction and semantic models of justification logic in the proof assistant coq. Logic journal of the IGPL  \textbf{28}(6),  1077--1092 (2020)

\bibitem{artemov-fitting-book}
Artemov, S.N., Fitting, M.: Justification Logic: Reasoning with Reasons. Cambridge University Press (2019)

\bibitem{benthem-12}
van Benthem, J., Fern\'andez-Duque, D., Pacuit, E.: Evidence logic: A new look at neighborhood structures (2012)

\bibitem{vBDE14}
van Benthem, J., Fern\'ndez-Duque, D., Pacuit, E.: Evidence and plausibility in neighborhood structures. Annals of pure and applied logic  \textbf{165}(1),  106--133 (2014)

\bibitem{from-21}
From, A.H.: Formalized soundness and completeness of epistemic logic. In: WoLLIC (2021)

\bibitem{goranko-21}
Goranko, V.: How deonic logic ought to be: towards a many-sorted framework for normative reasoning. In: DEON (2021)

\bibitem{liu-lorini-17}
Liu, F., Lorini, E.: Reasoning about belief, evidence and trust in a multi-agent setting. In: International Conference on Principles and Practice of Multi-Agent Systems (2017)

\bibitem{Lof84}
Martin-L\"{o}f, P.: Intuitionistic Type Theory. Bibliopolis, Naples (1984)

\bibitem{michaelis-nipkow-17}
Michaelis, J., Nipkow, T.: Formalized proof systems for propositional logic. In: TYPES (2017)

\bibitem{SBOM22}
NIST: Software security in supply chains: Software bill of materials ({SBOM}) (2022), \url{https://www.nist.gov/itl/executive-order-14028-improving-nations-cybersecurity/software-security-supply-chains-software-1}

\bibitem{PeC22}
Perera, J., Silva, C.C.: Organics case study: Use case briefs (2022), \url{https://veracity.wgtn.ac.nz/wp-content/uploads/2023/04/Organic_Supply_Chain__Use_Case_Briefs_Report.pdf}

\bibitem{Ree24}
Reeves, S.: A {L}ogic for {V}eracity (2024), \url{https://doi.org/10.48550/arXiv.2302.06164}

\bibitem{renne-12}
Renne, B.: Multi-agent justification logic: communication and evidence elimination. Synthese  (2012)

\bibitem{roennedal-18}
R\"onnedal, D.: Doxastic logic: a new approach. Journal of Applied Non-Classical Logics  (2018)

\bibitem{taglia-22}
Tagliaferri, M., Aldini, A.: From belief to trust: A quantitative framework based on modal logic. Journal of Logic and Computation  (2022)

\bibitem{coq}
{The Coq Development Team}: The {C}oq {P}roof {A}ssistant  (Jun 2023). \doi{10.5281/zenodo.1003420}, \url{https://coq.inria.fr}

\end{thebibliography}

\section*{Appendices}

\appendix

\section{The Logic}\label{sec:thelogic}

\subsection{Logical Rules}

\subsubsection{Preliminaries}

These purely logical rules do not change trust, so we have left the actor annotations off, since they will be the same for all witnesses mentioned. 


We have also, usually, omitted premises of the form $A \emph{ is a veracity claim}$ where this is clearly needed to make the rule well-formed. We have also omitted assumptions which appear unchanged in both the premises and the conclusion of a rule. So, for example:
$$
\begin{prooftree}
x \in A \vdash b(x) \in B
\justifies
\lambda b \in A \rightarrow B
\using
{\rightarrow^+}
\end{prooftree}
$$

is the abbreviated form of the full form:
$$
\begin{prooftree}
\Gamma \vdash A \emph{ is a veracity claim} \quad \Delta \vdash B \emph{ is a veracity claim} \quad \Theta, x \in A \vdash b(x) \in B
\justifies
\Gamma, \Delta, \Theta \vdash \lambda b \in A \rightarrow B
\using
{\rightarrow^+}
\end{prooftree}
$$

\subsubsection{The Rules}
$$
\begin{prooftree}
A \emph{ a veracity claim}
\justifies
x \in A \vdash x \in A
\using
{assume}
\end{prooftree}
\qquad
\begin{prooftree}
a \in A
\justifies
A \emph{ a veracity claim}
\using
{claim}
\end{prooftree}
$$

$$
\begin{prooftree}
a \in \bot
\justifies
a \in A
\using
{\bot^-}
\end{prooftree}
$$

$$
\begin{prooftree}
a \in A
\justifies
i(a) \in A \lor B
\using
{\lor^+ 1}
\end{prooftree}
\quad
\begin{prooftree}
b \in B
\justifies
j(b) \in A \lor B
\using
{\lor^+ 2}
\end{prooftree}
$$

$$ 
\begin{prooftree}
c \in A \lor B \quad x \in A \vdash d(x) \in C \quad y \in B \vdash e(y) \in C
\justifies
cases(c, d, e) \in C
\using
{\lor^-}
\end{prooftree}
$$

$$
\begin{prooftree}
c \in A \quad x \in A \vdash d(x) \in C \quad y \in B \vdash e(y) \in C
\justifies
cases(i(c), d, e = d(c) \in C
\using
{\lor^=1}
\end{prooftree}
$$

$$
\begin{prooftree}
c \in B \quad x \in A \vdash d(x) \in C \quad y \in B \vdash e(y) \in C
\justifies
cases(j(c), d, e) = e(c) \in C
\using
{\lor^=2}
\end{prooftree}
$$

$$
\begin{prooftree}
a_x \in A \quad b_y \in B
\justifies
(a,b) \in A \land B
\using
{\land^+}
\end{prooftree}
\quad
\begin{prooftree}
c \in A \land B \quad x \in A, y \in B \vdash d(x,y) \in C
\justifies
split(c, d) \in C
\using
{\land^-}
\end{prooftree}
$$

$$
\begin{prooftree}
a \in A \quad b \in B \quad x \in A, y \in B \vdash d(x,y) \in C
\justifies
split((a,b), d = d(a,b) \in C
\using
{\land^=}
\end{prooftree}
$$

$$
\begin{prooftree}
x_z \in A \vdash b(x) \in B
\justifies
\lambda b \in A \rightarrow B
\using
{\rightarrow^+}
\end{prooftree}
\quad
\begin{prooftree}
c \in A \rightarrow B \quad a \in A
\justifies
app(c, a) \in B
\using
{\rightarrow^-}
\end{prooftree}
$$

$$
\begin{prooftree}
a \in A \quad x \in A \vdash b(x) \in B
\justifies
app(\lambda b, a) = b(a) \in B
\using
{\rightarrow^=}
\end{prooftree}
$$


Note: $\lambda b$ is to be understood as $(\lambda b)$ but we leave parentheses off for brevity, and anyhow this is always clear from the context.

The equality rules in the above set are, as can be seen, nicely thought of a computation rules: rules which show us how to go from the non-canonical forms of expressions for witnesses that we get from elimination rules towards the canonical forms of expressions for witnesses that we get from introduction rules. As elsewhere, we view the canonical forms as ``the value" of an expression. 


\section{Families of Claims and Proof of Specialised $\lor$-elimination Rules (from Section \ref{sec:choice})}\label{app:choice}

In order to derive the simplified elimination rules in the paper we have to generalise the veracity claims (the propositions, $C$ is what follows) to make them \emph{families of sets} of claims. As usual with a family of sets, there is an index which comes from another set over which the family members range. 

This rule is:

$$
\begin{prooftree}
c \in A \lor B \quad z \in A \lor B \vdash C(z) \quad x \in A \vdash d(i(x)) \in C(c) \quad y \in B \vdash e(j(y)) \in C(c)
\justifies
cases(c, d, e) \in C(c)
\using
{\lor^-}
\end{prooftree}
$$
along with some equality rules for ``computing" with $cases$. Like:
\begin{align*}
& cases(i(a), d, e) = d(a) \\
& cases(j(b),d, e) = e(b)    
\end{align*}

In the example that follows $C(c)$, where $c \in A \lor B$, in the above rule is specialised (for this example) and is the family $\{A_{i(a)}, B_{j(b)}\}$, i.e. $C(c)$ is to be thought of as, more usually written, $\{C_y\}_{y \in A \lor B}$ for appropriate $C_y$s. An alternative way to say what $C(c)$ is might be to write
\begin{align*}
    C(i(a)) = A \\
    C(j(b)) = B
\end{align*}

Then rule $\lor^- 1$ above is derived as:
$$
\begin{prooftree}
i(a) \in A \lor B \quad x \in A \vdash x \in A \quad y \in B \vdash y \in B
\justifies
cases(i(a), (x)x, (y)y) \in A \lor B
\using
{\lor^-}
\end{prooftree}
$$

\noindent which simplifies, given the computation rules for $cases$ and the trivially holding second and third premises, to the rule given previously for $\lor^- 1$.

\section{Examples}

In this section we give simple examples of the logic in action.

\subsection{Stars v. Chains}

As a small example of how we can use the formalism of trust relations developed so far in characterising simple differences we consider a typical supply chain from a trust point-of-view and contrast it with a star supply ``chain" and see how the formalism shows the difference between them (an obvious difference, so this is a check that the formalism correctly characterises this, rather than a hard test for the formalism to solve).

The setting is a supply chain for some product (perhaps grain, perhaps grapes etc.) where producers, carriers, warehousers, wholesalers and retailers are all involved as actors as the product moves from countryside to consumer. We might picture this chain as

\vspace{.5cm}
\begin{center}
\begin{tikzpicture}
\usetikzlibrary {graphs}
\tikz
  \graph { 1/$p$ -- 2/$q$ -- 3/$r$ -- 4/$s$ -- 5/$t$ -- 6/$u$ };
\end{tikzpicture}
\end{center}

where the edges denote that movement of goods is possible between the actors $p$, $q$, ... (at the nodes). So, $p$ can pass goods to $q$ and vice versa, and so on.

The story behind this is: $p$ supplies good to storer $q$, who tells distributor $r$ that the goods are available and $r$ tells transporter $s$ that delivery can go ahead, so $s$ collects goods and delivers to retailer $t$ who finally supplies to customer $u$.

Now think about how information works in the chain. There is no reason it should follow the same path as the goods. In the usual supply chain it often does. So $u$ has no access to information that $p$ has, but only to information that $t$ has. This leads to inefficiencies in the chain's working: things might be more efficient if $u$ could see (some) information that $q$ has because it would increase veracity (trust, truth-telling etc.) and therefore the ease of working in the chain. A real example of this is where $p$ claims to have the means to pay for storage of the goods from where $u$ can collect them, but actually doesn't, leading to $u$ arriving at the storage to find a bill for their release. If $u$ could see the information about $p$ directly then they might be able to see that $p$ has paid (or not) the storage bill.

So, though the goods might move along the chain, it would be more efficient (due to better veracity) if information was arranged in a star where all participants have access to a pool of information:

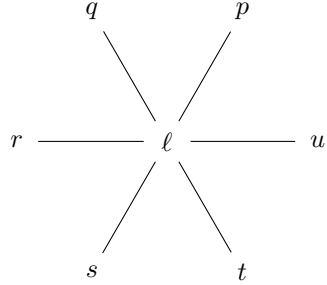
\vspace{.5cm}
\begin{center}
\begin{tikzpicture}
    \node[circle] at (360:0mm) (center) {$\ell$};
    
        \node[circle] at ({1*360/6}:2cm) (n1) {$p$};
        \node[circle] at ({2*360/6}:2cm) (n2) {$q$};
        \node[circle] at ({3*360/6}:2cm) (n3) {$r$};
        \node[circle] at ({4*360/6}:2cm) (n4) {$s$};
        \node[circle] at ({5*360/6}:2cm) (n5) {$t$};
        \node[circle] at ({6*360/6}:2cm) (n6) {$u$};
        \draw (center)--(n1);
        \draw (center)--(n2);
        \draw (center)--(n3);
        \draw (center)--(n4);
        \draw (center)--(n5);
        \draw (center)--(n6);
    
\end{tikzpicture}
\end{center}

We can formalise this to see the benefit. Let there be a (new) trust relation $S$ over the actors, weighted so that 
$$
pS_xq, qS_yr, rS_zs, sS_at, tS_bu
$$
for the chain. To take into account the star we can define $S$ so that
$$
pS_{1.0}\ell, qS_{1.0}\ell, rS_{1.0}\ell, sS_{1.0}\ell, tS_{1.0}\ell, uS_{1.0}\ell
$$
which records that each node completely trusts the central actor $\ell$ to keep a true record. (Perhaps ``$\ell$" is for ``ledger"....?)

$p$ needs to trust that $t$ will pay for the goods (via intermediate actors), and using the logic we can say that $pT_{x.y.z.a}t$. So, we can see that, unless there is complete trust ($x = y = z = a = 1$) between \emph{all} the actors then trust steadily decreases as information (about the payment being made) flows along the chain. In the case of the star, the very worst that can happen is that $t$ is untrustworthy to some degree (i.e. they don't always put true information into the ledger). If $t$ is trusted with a weight $c\%$ then $lS_{c}t$ and so $pS_{c}t$. 

This is a very simple case for illustration, but we can see that we now have a formal basis for calculating and discussing degrees of trust. In this case it is clear that the star will be better for $p$ as long as $c \geq x.y.z.a$.

In general it is clear that the longer the chain, the less trust there is. With the star, the trust level is constant (pairwise).

\subsection{Simple First Examples from some Case Studies}
\label{app:caseStudyExamples}

A typical (though very small) example of a derivation (proof tree) for a simple supply chain. 

We have:

\begin{itemize}
    \item $C_1$ is ``the fertilizer has these ingredients" (which might be further broken down into:
    \begin{itemize}
        \item $C_{11}$ this is the list of organic ingredients
        \item $C_{12}$ this is the list of non-organic ingredients
    \end{itemize})
    \item $C_2$ is ``the spreadsheet describes the ingredients of the fertilizer" (with similar components to $C_{11}$ and $C_{12}$ above)
    \item $C_3$ is ``the ingredients are all certified" (with component claims for each ingredient, perhaps)
    \item and so on...
\end{itemize}

Then we have a series of witnesses that support these claims. For example we might have $x^P \in C_1$ where $x^P$ is the evidence that Penelope uses (believes, cites, quotes) to support the claim that the the fertilizer has the ingredients as indicated. Similarly, $y^P \in C_2$ is a judgement where $y^P$ is the evidence that Penelope uses (believes, quotes, cites) to show that the spreadsheet is a true record of the ingredients. 
And finally, similarly for $z^P \in C_3$.


Then, an abbreviated form of the case study claim ``All things are listed and certified" comes out as
$$
C_1 \land C_2 \land C_3
$$

and a proof tree would be   

$$
\begin{prooftree}
\[\[C_1\ a\ veracity\ claim \justifies x^P \in C_1 \vdash x^P \in C_1 \using {assume}\] \quad \[C_2\ a\ veracity\ claim \justifies y^P \in C_2 \vdash y^P \in C_2 \using {assume}\]
\justifies
x^P \in C_1, y^P \in C_2 \vdash (x, y)^P \in C_1 \land C_2 
\using
{\land^+} \] \quad \[C_3\ a\ veracity\ claim \justifies z^P \in C_3 \vdash z^P \in C_3 \using {assume}\]
\justifies
x^P \in C_1, y^P \in C_2, z^P \in C_3 \vdash ((x, y),z)^P \in C_1 \land C_2 \land C_3
\using
{\land^+}
\end{prooftree}
$$

Note how the assumptions accumulate on the left of the turnstile, so we never lose assumptions (something that in a large example may well be a problem if keeping track of things by conventional means, perhaps).


We can give an example of another feature using the above example as a starting point. That is, we can show how a process (or function) for constructing evidence for complex claims can be built.

We can abstract on $x, y$ and $z$ above (the witnesses) to get the modified tree

$$
\begin{prooftree}
\[
\[C_1\ a\ veracity\ claim \justifies x^P \in C_1 \vdash x^P \in C_1 \using {assume}\] \quad \[C_2\ a\ veracity\ claim \justifies y^P \in C_2 \vdash y^P \in C_2 \using {assume}\]
\justifies
x^P \in C_1, y^P \in C_2 \vdash (x, y)^P \in C_1 \land C_2 
\using
{\land^+} \] \quad \[C_3\ a\ veracity\ claim \justifies z^P \in C_3 \vdash z^P \in C_3 \using {assume}\]
\justifies
\[x^P \in C_1, y^P \in C_2, z^P \in C_3 \vdash ((x, y),z)^P \in C_1 \land C_2 \land C_3
\using {\rightarrow^+}
\justifies
\[y^P \in C_2, z^P \in C_3 \vdash \lambda (x^P) ((x, y),z)^P \in C_1 \rightarrow (C_1 \land C_2 \land C_3)
\using {\rightarrow^+}
\justifies
\[z^P \in C_3 \vdash \lambda (y^P)(x^P) ((x, y),z)^P \in C_2 \rightarrow (C_1 \rightarrow (C_1 \land C_2 \land C_3)) 
\using {\rightarrow^+}
\justifies
\lambda (z^P)(y^P)(x^P) ((x, y),z)^P \in C_3 \rightarrow (C_2 \rightarrow (C_1 \rightarrow (C_1 \land C_2 \land C_3)))\] 
 \] \]  \using {\land^+}
\end{prooftree}
$$


We can see we have constructed a function which, given appropriate witnesses, constructs for us a witness to  (the conjunction which is) the required claim.

\subsection{Another Example---the process has been followed to justify a particular fact}


Assuming that Peter's status in the system is ``Completed", what supports the claim (how do we know the process that justifies the status has been followed) that this is a correct status?

The claim in the example rests on 13 steps. In order to make this a bit more tractable (without mechanisation) we'll consider just steps 3, 5, 6, 10 and 12. We assume here that we can simply check that the system  has Peter's status as ``Completed"; here we are interested in the claim that the process ending with this being that case has been followed. So, using $L_3,... L_{12}$ to stand for the relevant claims, we're looking for the evidence (the witness) for

$$
L_3 \rightarrow (L_5 \land L_6) \rightarrow L_{10} \rightarrow L_{12}
$$

We can build the following proof tree (where we've left out the leaves of the tree which simply make the assumptions that the $L$s are veracity claims, and that Peter is the actor in all cases):


$$
\begin{prooftree}
z \in L_{10}, y \in (L_5 \land L_6), x \in L_3 \vdash \ell \in L_{12}
\using {\rightarrow^+}
\justifies
\[y \in (L_5 \land L_6), x \in L_3 \vdash \lambda (z) \ell \in L_{10} \rightarrow L_{12}
\using {\rightarrow^+}
\justifies
\[x \in L_3 \vdash \lambda (y)(z) \ell  \in (L_5 \land L_6) \rightarrow L_{10} \rightarrow L_{12} 
\using {\rightarrow^+}
\justifies
\lambda (x)(y)(z) \ell \in L_3 \rightarrow (L_5 \land L_6) \rightarrow L_{10} \rightarrow L_{12}\] 
 \] \using {\rightarrow^+}
\end{prooftree}
$$

\section{Natural Language Rendering of the Proof from \Cref{list:processExample}, and example \Cref{app:caseStudyExamples} above}
\label{app:naturalLanguage}
\begin{itemize}
  \item (claim 3 implies (claim 2 implies (claim 1 implies ((claim 1 and claim 2) and claim 3)))) is supported by $\lambda(z)(\lambda(y)(\lambda(x)(((x, y), z))))$ which Penelope uses, because
    \begin{itemize}
    \item Assuming claim 3 is supported by $z$ which Penelope uses then (claim 2 implies (claim 1 implies ((claim 1 and claim 2) and claim 3))) is supported by $\lambda(y)(\lambda(x)(((x, y), z)))$ which Penelope uses, because
      \begin{itemize}
      \item Assuming claim 2 is supported by $y$ which Penelope uses, and claim 3 is supported by $z$ which Penelope uses then (claim 1 implies ((claim 1 and claim 2) and claim 3)) is supported by $\lambda(x)(((x, y), z))$ which Penelope uses, because
        \begin{itemize}
        \item Assuming claim 1 is supported by $x$ which Penelope uses, claim 2 is supported by $y$ which Penelope uses, and claim 3 is supported by $z$ which Penelope uses then ((claim 1 and claim 2) and claim 3) is supported by $((x, y), z)$ which Penelope uses, because
          \begin{itemize}
          \item Assuming claim 1 is supported by $x$ which Penelope uses, and claim 2 is supported by $y$ which Penelope uses then (claim 1 and claim 2) is supported by $(x, y)$ which Penelope uses, because
            \begin{itemize}
            \item Assuming claim 1 is supported by $x$ which Penelope uses then claim 1 is supported by $x$ which Penelope uses, because
              \begin{itemize}
              \item claim 1 is a veracity claim.
              \end{itemize}
            \item by assumption.
            \item Assuming claim 2 is supported by $y$ which Penelope uses then claim 2 is supported by $y$ which Penelope uses, because
              \begin{itemize}
              \item claim 2 is a veracity claim.
              \end{itemize}
            \item by assumption.
            \end{itemize}
          \item by a logical rule for 'and'.
          \item Assuming claim 3 is supported by $z$ which Penelope uses then claim 3 is supported by $z$ which Penelope uses, because
            \begin{itemize}
            \item claim 3 is a veracity claim.
            \end{itemize}
          \item by assumption.
          \end{itemize}
        \item by a logical rule for 'and'.
        \end{itemize}
      \item by a logical rule for implication.
      \end{itemize}
    \item by a logical rule for implication.
    \end{itemize}
  \item by a logical rule for implication.
  \end{itemize}

\section{Automatically Generated Proofs of $C \wedge C \wedge C \wedge C$}
\label{app:automation}

\begin{scprooftree}{1}\AxiomC{$ C \wedge C \textit{ is a veracity claim} $} \RightLabel{ $ assume $}\UnaryInfC{$ e^{a_{1}} \in C \wedge C \vdash_{} e^{a_{1}} \in C \wedge C $}\AxiomC{$ C \wedge C \textit{ is a veracity claim} $} \RightLabel{ $ assume $}\UnaryInfC{$ e^{a_{1}} \in C \wedge C \vdash_{} e^{a_{1}} \in C \wedge C $} \RightLabel{ $ \wedge^{+} $} \BinaryInfC{$ e^{a_{1}} \in C \wedge C \vdash_{} (e, e)^{a_{1}} \in C \wedge C \wedge C \wedge C $}\end{scprooftree}

\begin{scprooftree}{0.8}\AxiomC{$ C \textit{ is a veracity claim} $} \RightLabel{ $ assume $}\UnaryInfC{$ e^{a_{1}} \in C \vdash_{} e^{a_{1}} \in C $}\AxiomC{$ C \textit{ is a veracity claim} $} \RightLabel{ $ assume $}\UnaryInfC{$ e^{a_{1}} \in C \vdash_{} e^{a_{1}} \in C $} \RightLabel{ $ \wedge^{+} $} \BinaryInfC{$ e^{a_{1}} \in C \vdash_{} (e, e)^{a_{1}} \in C \wedge C $}\AxiomC{$ C \wedge C \textit{ is a veracity claim} $} \RightLabel{ $ assume $}\UnaryInfC{$ e^{a_{1}} \in C \wedge C \vdash_{} e^{a_{1}} \in C \wedge C $} \RightLabel{ $ \wedge^{+} $} \BinaryInfC{$ e^{a_{1}} \in C, e^{a_{1}} \in C \wedge C \vdash_{} ((e, e), e)^{a_{1}} \in C \wedge C \wedge C \wedge C $}\end{scprooftree}

\begin{scprooftree}{0.8}\AxiomC{$ C \wedge C \textit{ is a veracity claim} $} \RightLabel{ $ assume $}\UnaryInfC{$ e^{a_{1}} \in C \wedge C \vdash_{} e^{a_{1}} \in C \wedge C $}\AxiomC{$ C \textit{ is a veracity claim} $} \RightLabel{ $ assume $}\UnaryInfC{$ e^{a_{1}} \in C \vdash_{} e^{a_{1}} \in C $}\AxiomC{$ C \textit{ is a veracity claim} $} \RightLabel{ $ assume $}\UnaryInfC{$ e^{a_{1}} \in C \vdash_{} e^{a_{1}} \in C $} \RightLabel{ $ \wedge^{+} $} \BinaryInfC{$ e^{a_{1}} \in C \vdash_{} (e, e)^{a_{1}} \in C \wedge C $} \RightLabel{ $ \wedge^{+} $} \BinaryInfC{$ e^{a_{1}} \in C \wedge C, e^{a_{1}} \in C \vdash_{} (e, (e, e))^{a_{1}} \in C \wedge C \wedge C \wedge C $}\end{scprooftree}

\begin{scprooftree}{0.63}\AxiomC{$ C \textit{ is a veracity claim} $} \RightLabel{ $ assume $}\UnaryInfC{$ e^{a_{1}} \in C \vdash_{} e^{a_{1}} \in C $}\AxiomC{$ C \textit{ is a veracity claim} $} \RightLabel{ $ assume $}\UnaryInfC{$ e^{a_{1}} \in C \vdash_{} e^{a_{1}} \in C $} \RightLabel{ $ \wedge^{+} $} \BinaryInfC{$ e^{a_{1}} \in C \vdash_{} (e, e)^{a_{1}} \in C \wedge C $}\AxiomC{$ C \textit{ is a veracity claim} $} \RightLabel{ $ assume $}\UnaryInfC{$ e^{a_{1}} \in C \vdash_{} e^{a_{1}} \in C $}\AxiomC{$ C \textit{ is a veracity claim} $} \RightLabel{ $ assume $}\UnaryInfC{$ e^{a_{1}} \in C \vdash_{} e^{a_{1}} \in C $} \RightLabel{ $ \wedge^{+} $} \BinaryInfC{$ e^{a_{1}} \in C \vdash_{} (e, e)^{a_{1}} \in C \wedge C $} \RightLabel{ $ \wedge^{+} $} \BinaryInfC{$ e^{a_{1}} \in C \vdash_{} ((e, e), (e, e))^{a_{1}} \in C \wedge C \wedge C \wedge C $}\end{scprooftree}

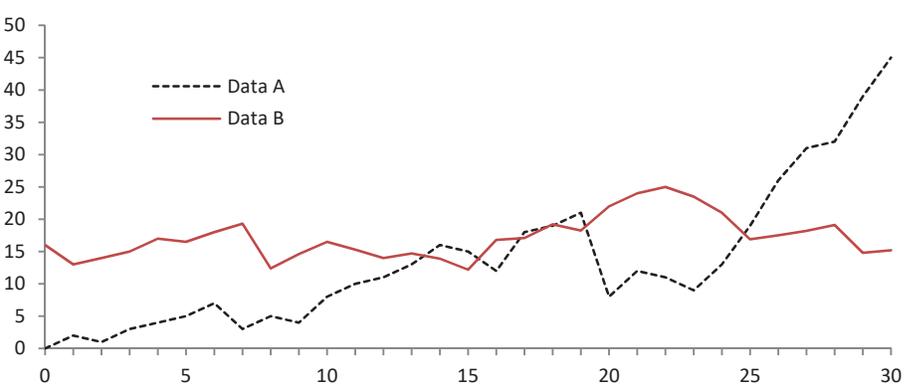
\end{document}